\documentclass[runningheads]{llncs}

\usepackage[T1]{fontenc}
\usepackage[utf8]{inputenc}
\usepackage{./sty/mrstylefile}

\usepackage{color}

\usepackage{./sty/tikzit}
\usetikzlibrary{shapes}
\usetikzlibrary{arrows}

\tikzstyle{vertex}=[inner sep=0mm, outer sep=1mm, shape=circle, fill=none]

\tikzstyle{reaction}=[->, draw=black]
\tikzstyle{activation}=[->, draw={rgb,255: red,25; green,154; blue,52}, thick]
\tikzstyle{inhibition}=[{-|}, draw={rgb,255: red,205; green,30; blue,53}, thick]
\tikzstyle{transition}=[->, draw={rgb,255: red,25; green,47; blue,154}]

\graphicspath{ {./figures/} }

\begin{document}

\title{Modulation-Reaction Networks}

\titlerunning{Modulation-Reaction Networks}

\author{Leo Lobski\thanks{Both authors contributed equally to this work.}\inst{1}\orcidID{0000-0002-0260-0240} \and
Yo\`av Montacute$^\star$\inst{2}\orcidID{0000-0001-9814-7323}}

\authorrunning{L.~Lobski and Y.~Montacute}

\institute{University College London, London, UK \\ \email{leo.lobski.21@ucl.ac.uk}
\and
National Institute of Informatics, Tokyo, Japan \\ \email{montacute@nii.ac.jp}}

\maketitle

\begin{abstract}
Biochemical systems involve both the flow of matter, in which entities transform into one another via reactions, and the flow of information, in which entities regulate which reactions may occur.
Boolean networks capture the latter; reaction networks capture the former. 
Yet no unified qualitative formalism treats regulated reactions as its principal objects of study, despite their prominence in standards such as the Systems Biology Graphical Notation Process Description (SBGN-PD) language.
We introduce modulation-reaction networks (MR-networks), a mathematical framework in which entities modulate reactions through activations and inhibitions, and study their synchronous Boolean semantics.
To reason about MR-networks we develop Modulation-Reaction Logic (MRL), a hybrid modal $\mu$-calculus whose modalities reason about the structure of the network and whose fixed-point operators capture temporal evolution of the computation.
We establish a collection of validities, including a complete characterisation of the one-step update rule, and demonstrate the expressive power of MRL by formalising properties of biological interest such as reachability, sustained production, and presence of attractors. 
We show that MRL admits model-checking via an evaluation game, and introduce a bisimulation relation for MR-networks, which is proved to be invariant for all MRL-formulas.
As a step towards a biologically more realistic computational model, we sketch the asynchronous semantics of MR-networks, and outline how the developments for the synchronous case transfer to the study of the asynchronous one.

\keywords{Boolean network \and Reaction network \and SBGN-PD \and MR-network \and Modal logic \and $\mu$-calculus \and Model checking}
\end{abstract}

\section{Introduction}

\begin{figure}[h]
\centering
\vspace{-1em}
\scalebox{1}{\begin{tikzpicture}
	\begin{pgfonlayer}{nodelayer}
		\node [style=vertex] (0) at (-7, 1) {$x$};
		\node [style=vertex] (1) at (-2.5, 1) {$y$};
		\node [style=vertex] (2) at (-4.75, -1) {$z$};
		\node [style=none] (3) at (-8.5, 1) {\textbf{A}};
		\node [style=vertex] (4) at (2, 1) {$i_1$};
		\node [style=vertex] (5) at (2, -1) {$i_2$};
		\node [style=vertex] (6) at (5, 1) {$x$};
		\node [style=vertex] (7) at (11, 1) {$o_1$};
		\node [style=vertex] (8) at (11, -1) {$o_2$};
		\node [style=vertex] (9) at (8, -1) {$y$};
		\node [style=none] (10) at (8, 1.25) {};
		\node [style=none] (11) at (8, 0.75) {};
		\node [style=none] (12) at (5, -0.75) {};
		\node [style=none] (13) at (0.5, 1) {\textbf{B}};
		\node [style=none] (14) at (3.5, 1.25) {};
		\node [style=none] (15) at (9.5, -1.25) {};
	\end{pgfonlayer}
	\begin{pgfonlayer}{edgelayer}
		\draw [style=activation] (1) to (0);
		\draw [style=activation, in=135, out=45, loop] (0) to ();
		\draw [style=inhibition] (2) to (0);
		\draw [style=inhibition, bend right] (2) to (1);
		\draw [style=inhibition, bend right] (1) to (2);
		\draw [style=reaction] (4) to (6);
		\draw [style=reaction] (6) to (7);
		\draw [style=reaction] (5) to (9);
		\draw [style=reaction] (9) to (8);
		\draw [style=activation] (6) to (12.center);
		\draw [style=activation] (9) to (11.center);
		\draw [style=inhibition, bend left=90] (6) to (10.center);
		\draw [style=activation, bend left=90, looseness=1.50] (4) to (14.center);
		\draw [style=activation, bend right=90, looseness=1.50] (9) to (15.center);
	\end{pgfonlayer}
\end{tikzpicture}}
\vspace{-1em}
\caption{\textbf{A}: An example of a regulatory graph taken from~\cite[Fig.~1]{cury26}. \textbf{B}: An example MR-network, whose structure is taken from~\cite[Fig.~9]{voit13}.\label{fig:reg-graph}}
\AltTextCMSB{A regulatory graph and an MR-network}
\end{figure}
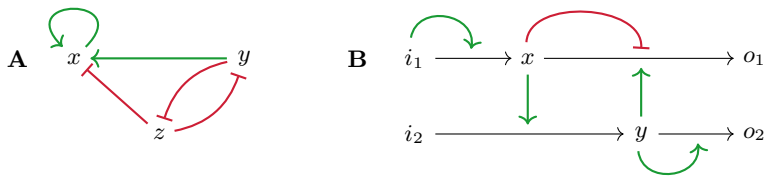

{\em Boolean networks} are a widely used and studied model for regulatory networks in biology and related sciences~\cite{kauffman1969,thomas1973,thieffry1999,bornholdt2008,wang2012,boolean-networks2022}. A Boolean network features a {\em regulatory graph}~\cite{compositional-account2024}, which consists of a set of vertices that represent the entities (e.g.~genes) which regulate each other via two kinds of edges: {\em activations} and {\em inhibitions}.
We give an example of a regulatory graph in Fig.~\ref{fig:reg-graph}(A). The {\em state} of a Boolean network is simply an assignment of Boolean truth values to each vertex of the regulatory graph. The interpretation is that an entity (e.g.~signal, molecule or a phenotype) is either present (active) or absent (inactive). A model of computation provides the rules that are used to transition from one state to another: this takes the form of a Boolean function for each node, together with conditions under which the truth value of the state is updated by applying the function.
We note that in the usual presentation, the Boolean functions are taken to be part of the data of the Boolean network, rather than part of the semantics.
We choose to view them as part of the semantics, while the syntax is provided by the regulatory graph. This allows interpreting the regulatory graphs under other dynamics, including quantitative ones.
A Boolean network can be thought to represent the {\em logical} or {\em information} flow between the entities: instead of capturing transformations between entities, it tells how they activate or inhibit each other.

{\em Reaction networks}, on the other hand, represent transformations between entities, by describing how reactants transform into products~\cite{inferring-reaction-systems2015,abstract-simulation2022}. At the bare minimum, a reaction network consists of a graph, whose edge indicates a reaction between chemical or biological entities. A reaction network, therefore, represents {\em matter} or {\em entity} flow. Depending on the level of abstraction, a reaction network may include enough information to describe a dynamical system, or it may be purely qualitative, describing {\em possibilities} of transformations rather than rates of change. Qualitative networks can capture essential features of interest by abstracting away the details, thereby reducing computational complexity. Since we are interested in combining reaction networks with Boolean networks, we mostly focus on qualitative semantics.
A well-studied qualitative semantics of reaction networks is given by the Boolean semantics of BIOCHAM~\cite{biocham2005,biocham2006}, where, akin to Boolean networks, the state of a network is represented by an assignment of Boolean truth values, but crucially to both all the entities {\em and} all the reactions.
The interpretation is that each entity is either present or absent, and that each reaction is either {\em active} or {\em inactive}.
The transitions of BIOCHAM take place asynchronously, resulting in nondeterministic semantics.

\begin{figure}[h]
\centering
\includegraphics[scale=0.6]{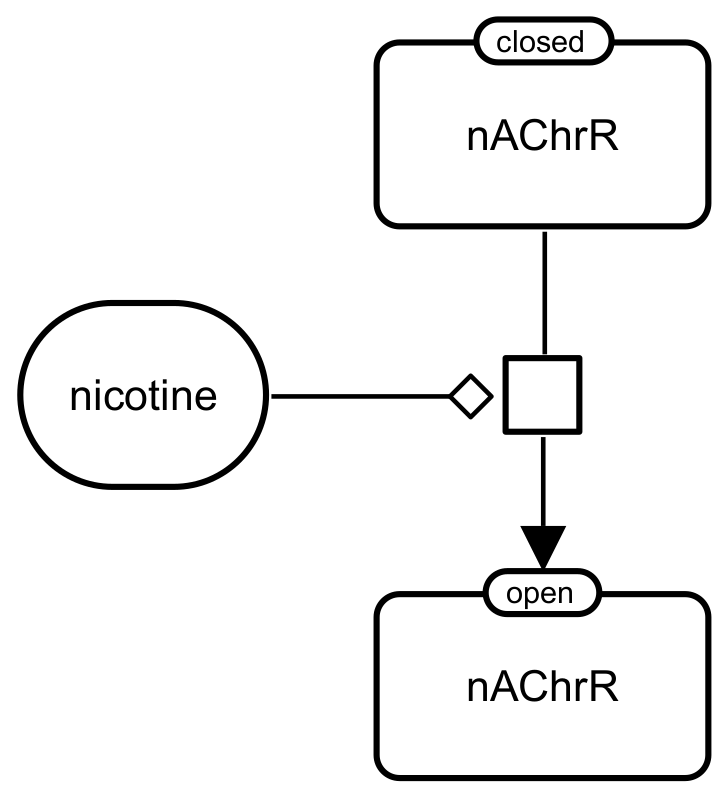}
\caption{An example of an SBGN-PD diagram with one process and one modulation, taken from~\cite[Fig.~2.42]{sbgn-pd19}.\label{fig:sbgn-nicotine}}
\AltTextCMSB{An SBGN-PD diagram}
\end{figure}

Any modelling of biochemical processes involves a mix of regulatory and process-like behaviour, and hence a mix of information and matter flow.
This is accounted for in the approaches to chemical reaction networks including inhibitors as part of reaction data~\cite{inferring-reaction-systems2015,influence-compared2018}.
For example, the diagram in Fig.~\ref{fig:sbgn-nicotine} represents modulation of nicotinic receptor opening by nicotine. 
Similarly, Fig.~\ref{fig:reg-graph}(B) depicts a system where $x$ activates the production of $y$ and inhibits its own degradation, while $y$ activates the degradation of $x$. 
Systems Biology Graphical Notation (SBGN) is a graphical notational standard for representing biological processes, featuring three levels of description, of which Process Description (SBGN-PD) captures both matter flow in the form of the {\em process nodes} and information flow in the form of the {\em modulations}. 
Fig.~\ref{fig:sbgn-nicotine} shows an example of a (rather simple) SBGN-PD diagram. 
The feature distinguishing SBGN-PD diagrams from both Boolean and reaction networks is that, instead of directly regulating other entities, the entities regulate the processes (via the modulations). This can be thought of as a refinement of a Boolean network: instead of describing the regulation relations between the entities, the precise mechanism via which the regulation happens is included~\cite{influence-compared2018,compositional-account2024}.

In this work, taking inspiration from SBGN-PD, we introduce {\em modulation-reaction networks} (MR-networks) as formal mathematical models of regulated reactions. 
While here we consider only two kinds of regulations, {\em activation} and {\em inhibition}, we expect this framework to generalise to MR-networks with more diverse regulatory components, ultimately yielding a mathematical theory of all the modulations allowed in SBGN-PD and more broadly in biological modelling.

As in reaction networks, we represent the matter flow as a graph. 
The graph, however, varies with time, as the edges can appear and disappear as they are being activated or inhibited by the entities, much like in Boolean networks. 
Fig.~\ref{fig:reg-graph}(B) shows an example of an MR-network.
We study Boolean synchronous semantics for the MR-networks in detail, and give an overview of asynchronous BIOCHAM style semantics and of discrete qualitative semantics.
In order to reason with MR-networks, we introduce a formal language and a logic $\mathbf{MRL}$ that is sound with respect to the computational semantics.
From the logical point of view, we treat MR-networks as a sequence of labelled transition systems: the modalities control which transitions affect the truth value of a formula. 
We identify formulas that capture biologically interesting properties of the MR-networks.
While the developments of this paper mostly pertain to the synchronous update rule introduced in Section~\ref{sec:mr-networks}, we emphasise that the formulas of $\mathbf{MRL}$ can be interpreted over the asynchronous semantics outlined in Section~\ref{sec:other-models}. This, in particular, applies to the formulas describing biologically relevant network properties of Section~\ref{sec:expressivity}. Asynchronous semantics is more interesting from the biological point of view: it is more realistic to assume that the transitions occur locally in their own time rather than following a global clock. 
The precise validities and network properties under the asynchronous semantics are the subject of a forthcoming paper.

\paragraph{Structure of the paper.} In Section~\ref{sec:mr-networks}, we define MR-networks and their synchronous semantics. Section~\ref{sec:formal-language} introduces the logic of MR-networks $\mathbf{MRL}$ and establishes some validities. In Section~\ref{sec:expressivity}, we further study the formulas of $\mathbf{MRL}$ and the network properties they correspond to, with particular emphasis on biologically relevant properties. Section~\ref{sec:bisimulation} introduces evaluation games and the bisimulation relation for MR-networks: the former is used for model-checking, the latter is used to obtain inexpressibility results. In Section~\ref{sec:other-models}, we give a translation from MR-networks to Boolean networks, and discuss alternative semantics, including the asynchronous semantics. Section~\ref{sec:conclusion} concludes and outlines directions for future work.

\section{MR-networks}\label{sec:mr-networks}
We introduce our main objects of study.
\begin{definition}[MR-network]
An {\em MR-network} $(V,R,M^\activation,M^\inhibition)$ consists of a finite set $V$ of {\em vertices}, a binary relation $R\sse V\times V$ whose pairs are called {\em reactions}, and binary relations $M^\activation,M^\inhibition\sse V\times R$ whose pairs are called {\em modulations}.
\end{definition}
We will often write $uRv$ to denote $(u,v)\in R$. If $uRv$, we say that $u$ is the {\em source} and $v$ is the {\em target} of the reaction. If $w M^\activation r$, we say that the vertex $w$ {\em activates} the reaction $r$, and if $w M^\inhibition r$, we say that $w$ {\em inhibits} $r$. In both cases, we say that $w$ {\em modulates} $r$.

\vspace{3pt}
\noindent\begin{minipage}{.77\textwidth}
\begin{example}\label{ex:mr-network-det}
On the right, we draw the MR-network with $V=\{x,y,z\}$, $R=\{(x,y)\}$, $M^\activation=\{(z,(x,y))\}$ and $M^\inhibition=\eset$. It formalises the SBGN-PD diagram of Fig.~\ref{fig:sbgn-nicotine}, once the modulation is interpreted as activation.
\end{example}
\end{minipage}%
\begin{minipage}{.23\textwidth}
\vspace{-15pt}
\begin{equation*}
\scalebox{1}{\begin{tikzpicture}
	\begin{pgfonlayer}{nodelayer}
		\node [style=vertex] (0) at (-2, 0) {$x$};
		\node [style=vertex] (1) at (2, 0) {$y$};
		\node [style=vertex] (2) at (0, 2) {$z$};
		\node [style=none] (3) at (0, 0.25) {};
	\end{pgfonlayer}
	\begin{pgfonlayer}{edgelayer}
		\draw [style=reaction] (0) to (1);
		\draw [style=activation] (2) to (3.center);
	\end{pgfonlayer}
\end{tikzpicture}
}
\end{equation*}
\end{minipage}
\vspace{3pt}

Next, we define the synchronous update rule of MR-networks, which will be the semantics studied for the majority of the paper; alternative semantics are discussed in Section~\ref{sec:other-models}. We write $\Z_2=\{0,1\}$ for the integers modulo $2$.
\begin{definition}[Valuation and state]
Let $(V,R,M^\activation,M^\inhibition)$ be an MR-network. A {\em valuation} is a function $f:V\rightarrow\Z_2$. A {\em state} is a pair $(f,g)$ of a valuation $f$ and a function $g:R\rightarrow\Z_2$.
\end{definition}

Given a valuation $f:V\rightarrow\Z_2$, it induces the function $g:R\rightarrow\Z_2$ as follows:
\begin{equation}\label{eq:reaction-update}
g(u,v) \coloneq f(u)\land\bigvee_{w\in V : M^\activation(w,(u,v))}f(w)\land\bigwedge_{w\in V : M^\inhibition(w,(u,v))}\neg f(w).
\end{equation}
Given a state $(f,g)$, it induces a valuation $f':V\rightarrow\Z_2$ as follows:
\begin{equation}\label{eq:vertex-update}
f'(u) \coloneq f(u)\lor\bigvee_{w\in V : (w,u)\in R}g(w,u).
\end{equation}
A {\em transition} is a pair of states, written as $(f,g)\longrightarrow (f',g')$ such that $f'$ is the valuation induced by $(f,g)$, and $g'$ is the function induced by $f'$.

\begin{definition}[Computation]
Let $(V,R,M^\activation,M^\inhibition)$ be an MR-network, and let $f_0:V\rightarrow\Z_2$ be a valuation. The {\em computation} starting from $f_0$ is the sequence of states $\left\{(f_i,g_i)\right\}_{i\in\N}$ such that $(f_0,g_0)$ is the state extending $f_0$, and for each $i\in\N$, the pair $\left(f_i,g_i\right)\longrightarrow \left(f_{i+1},g_{i+1}\right)$ is the state transition induced by $\left(f_i,g_i\right)$.
\end{definition}

\begin{example}\label{ex:mr-network2-det}
Consider the MR-network obtained from Example~\ref{ex:mr-network-det} by making $y$ an inhibitor of the reaction, i.e.~$M^\inhibition=\{(y,(x,y))\}$.
With the initial valuation $x,z\mapsto 1$ and $y\mapsto 0$, we obtain the computation below. Note that the reaction $(x,y)$ is inactive in the fixed-point, so that the production of $y$ is not sustained.
\begin{equation*}
\scalebox{1}{\begin{tikzpicture}
	\begin{pgfonlayer}{nodelayer}
		\node [style=vertex] (0) at (-11.25, -0.5) {$x_1$};
		\node [style=vertex] (1) at (-7.25, -0.5) {$y_0$};
		\node [style=vertex] (2) at (-9.75, 1.5) {$z_1$};
		\node [style=none] (3) at (-9.75, -0.25) {};
		\node [style=none] (4) at (-8.75, -0.25) {};
		\node [style=none] (5) at (-6.25, 0.5) {};
		\node [style=none] (6) at (-4.25, 0.5) {};
		\node [style=none] (7) at (-9.25, -0.75) {$_1$};
		\node [style=vertex] (8) at (-3.25, -0.5) {$x_1$};
		\node [style=vertex] (9) at (0.75, -0.5) {$y_1$};
		\node [style=vertex] (10) at (-1.75, 1.5) {$z_1$};
		\node [style=none] (11) at (-1.75, -0.25) {};
		\node [style=none] (12) at (-0.75, -0.25) {};
		\node [style=none] (13) at (-1.25, -0.75) {$_1$};
		\node [style=none] (14) at (1.75, 0.5) {};
		\node [style=none] (15) at (3.75, 0.5) {};
		\node [style=vertex] (16) at (4.75, -0.5) {$x_1$};
		\node [style=vertex] (17) at (8.75, -0.5) {$y_1$};
		\node [style=vertex] (18) at (6.25, 1.5) {$z_1$};
		\node [style=none] (19) at (6.25, -0.25) {};
		\node [style=none] (20) at (7.25, -0.25) {};
		\node [style=none] (21) at (6.75, -0.75) {$_0$};
		\node [style=none] (22) at (9.75, 0.5) {};
		\node [style=none] (23) at (6.75, -1.75) {};
	\end{pgfonlayer}
	\begin{pgfonlayer}{edgelayer}
		\draw [style=reaction] (0) to (1);
		\draw [style=activation] (2) to (3.center);
		\draw [style=inhibition, in=90, out=90, looseness=2.75] (1) to (4.center);
		\draw [style=transition] (5.center) to (6.center);
		\draw [style=reaction] (8) to (9);
		\draw [style=activation] (10) to (11.center);
		\draw [style=inhibition, in=90, out=90, looseness=2.75] (9) to (12.center);
		\draw [style=transition] (14.center) to (15.center);
		\draw [style=reaction] (16) to (17);
		\draw [style=activation] (18) to (19.center);
		\draw [style=inhibition, in=90, out=90, looseness=2.75] (17) to (20.center);
		\draw [style=transition, in=-60, out=-15, looseness=2.00] (22.center) to (23.center);
	\end{pgfonlayer}
\end{tikzpicture}}
\vspace{-3em}
\end{equation*}
\end{example}

\section{Formal language}\label{sec:formal-language}
We define the formal language of {\em Modulation-Reaction Logic} ($\mathbf{MRL}$). 
The syntax of $\mathbf{MRL}$ is defined recursively as follows:
$$ \varphi::= 0 \mid 1 \mid \neg \varphi \mid  \varphi\wedge\psi \mid \lozenge^+\varphi \mid \lozenge^-\varphi \mid X\varphi  \mid \mu p.\varphi\mid i \mid @_i \varphi,$$
where $i\in \mathbf{Nom}$; the set $\mathbf{Nom}$ is a finite set whose members are called \emph{nominals}.
In $\mu p.\varphi$, we require that $p$ occurs only positively in $\varphi$ (i.e.\ under an even number of negations).
We define the Boolean operations $\vee,\to $ and $\leftrightarrow$ in the usual way; we define $\top:= 0\vee 1$ and $\bot:= 0\wedge 1$;
we define the duals $\square^{(\cdot)}:=\neg \lozenge^{(\cdot)}\neg$, where ${(\cdot)}\in \{+,-\}$, and $\nu p.\varphi:=\neg\mu p.\neg\varphi[\neg p / p]$. 
We further abbreviate $F\varphi:=\mu p.(\varphi\vee Xp)$ and $G\varphi:=\neg F \neg \varphi$.

The constants $0$ and $1$ test the current activation value of a vertex.
The connectives $\neg$ and $\wedge$ are Boolean.
The \emph{network modalities} express movement on the graph: $\lozenge^{\!+}\varphi$ (resp.\ $\lozenge^{\!-}\varphi$) holds at a vertex if some successor (resp.\ predecessor) along an active reaction satisfies $\varphi$.
The \emph{computation modalities} express movement on the temporal axis: $X\varphi$ holds if $\varphi$ holds at the next computation step.
The derived modalities $F\varphi$ and $G\varphi$ hold if $\varphi$ holds at some (resp.\ every) future step.
The \emph{fixed-point operator} $\mu p.\varphi$ denotes the least property $p$ satisfying $\varphi$; its dual $\nu p.\varphi$ denotes the greatest such property. 
The \emph{hybrid operators} allow reference to specific vertices: a nominal $i$ holds at $x$ iff $x$ is the vertex named $i$, and $@_i\varphi$ asserts that $\varphi$ holds at the vertex named $i$.

\paragraph{Semantics.}
By an {\em MR-model} we mean a pair $(\mathcal M,f_0)$ of an MR-network $\mathcal M$ and a valuation $f_0$. An MR-model thus induces a sequence of states $\{(f_i,g_i)\}_{i\in\N}$, and hence a sequence of MR-models $\{\mathcal M_i\}_{i\in\N}$ defined by $\mathcal M_i\coloneq (\mathcal M,f_i)$.

The fixed-point operator $\mu p.\varphi$ binds the variable $p$; evaluating $\varphi$ during the fixed-point computation requires interpreting $p$ as a set of vertex-time pairs that changes at each approximation stage. To handle this, satisfaction is defined relative to an \emph{assignment} $\sigma : \mathbf{Var}\to\mathcal{P}(V\times\N)$, which maps each propositional variable to the set of pairs $(x,i)$ at which it is considered true. We write $\sigma[p\mapsto S]$ for the assignment that maps $p$ to $S$ and agrees with $\sigma$ otherwise.

In the clauses in Table~\ref{tab:satisfaction}, we assume a fixed MR-model $(\mathcal M,f_0)$ and an assignment $\sigma$; we refer to the whole chain of induced models $\mathcal M_i$ and states $(f_i,g_i)$. Given $x\in V$ we define the satisfaction relation $\models_\sigma$ recursively in Table~\ref{tab:satisfaction}. 
For the fixed-point modality, we note that the positivity condition on $p$ ensures that $\Phi^{\varphi,p}_\sigma$ is monotone, so the least fixed-point exists by the Knaster-Tarski theorem. 
On finite MR-networks with monotone activation, the valuation $\{f_i\}_{i\in\N}$ stabilises: the fixed-point is therefore reached in finitely many iterations of $\Phi^{\varphi,p}_\sigma$ and is thus computable.
A formula with no free propositional variables is called a \emph{sentence}; for sentences, the satisfaction relation does not depend on~$\sigma$, and we write simply $\mathcal{M}_i,x\models\varphi$.
\vspace{-1em}
\begin{table}
\fbox{%
\begin{minipage}{\textwidth}%
\vspace{-10pt}
\begin{flalign*}
&\textbf{Network modalities} &&\\
\mathcal{M}_i,x &\models_\sigma j  \iff f_i(x)=j, for j\in\mathbb{Z}_2 &&\\
\mathcal{M}_i,x &\models_\sigma p \iff (x,i)\in\sigma(p) &&\\
\mathcal{M}_i,x &\models_\sigma \neg\varphi  \iff  \mathcal{M}_i,x\not\models_\sigma \varphi &&\\
\mathcal{M}_i,x &\models_\sigma \varphi\wedge\psi \iff \mathcal{M}_i,x\models_\sigma \varphi \text{ and } \mathcal{M}_i,x \models_\sigma \psi &&\\
\mathcal{M}_i,x &\models_\sigma \lozenge\!^+\varphi\iff \exists y\in V \text{ s.t. } x R y, g_i(x,y)=1 \text{ and } \mathcal{M}_i,y\models_\sigma \varphi &&\\
\mathcal{M}_i,x &\models_\sigma \lozenge\!^-\varphi\iff \exists y\in V \text { s.t. } y R x, g_i(y,x)=1 \text{ and } \mathcal{M}_i,y\models_\sigma \varphi &&\\
&\textbf{Computation modalities} &&\\
\mathcal{M}_i,x &\models_\sigma X\varphi\iff \mathcal{M}_{i+1},x\models_\sigma\varphi &&\\
\mathcal{M}_i,x &\models_\sigma F\varphi\iff \mathcal{M}_{\ell},x\models_\sigma\varphi \text{ for some } \ell\geq i &&\\
&\textbf{Fixed-point modality} &&\\
\mathcal{M}_i,x &\models_\sigma \mu p.\varphi \iff (x,i)\in\bigcap\big\{S\subseteq V\times\N : \Phi^{\varphi,p}_\sigma(S)\subseteq S\big\}, &&\\
&\text{where } \Phi^{\varphi,p}_\sigma(S)=\{(x',i')\in V\times\N : \mathcal{M}_{i'},x'\models_{\sigma[p\mapsto S]}\varphi\} &&\\
&\textbf{Hybrid modalities} &&\\
\mathcal{M}_i,x &\models_\sigma y\iff x=y &&\\
\mathcal{M}_i,x &\models_\sigma @_y \varphi \iff \mathcal{M}_i,y\models_\sigma\varphi &&
\end{flalign*}
\end{minipage}}
\vspace{1pt}
\caption{Satisfaction of $\mathbf{MRL}$ formulas.\label{tab:satisfaction}}
\vspace{-3em}
\end{table}

\vspace{1em}
An MR-model $(\mathcal M,f_0)$ {\em models} a sentence $\varphi\in\mathbf{MRL}$ at time $i$, written $\mathcal M_i\models\varphi$, if for all $x\in V$ we have $\mathcal M_i,x\models\varphi$. 
A sentence $\varphi$ is \emph{valid in} an MR-network $\mathcal M$ if for all valuations $f_0$ we have $\mathcal{M}_0\models\varphi$. 
A sentence $\varphi$ is {\em valid} if it is valid in all MR-networks. We write $\mathcal M\models\varphi$ for validity in $\mathcal M$, and $\models\varphi$ for validity.
We write $\varphi\equiv\psi$ for logical equivalence, i.e.~iff $\varphi \leftrightarrow \psi$ is valid. 

\begin{proposition}\label{prop:validities}
    The following formulas are valid:
\begin{align*}
\customlabel{validity:persistence}{\text{(i)}}\ &1\to X1 &
\customlabel{validity:update}{\text{(ii)}}\ &X1\leftrightarrow (1\vee\lozenge^{\!-}\top) &
\customlabel{validity:convergence}{\text{(iii)}}\ &FG1\vee G0 &
\customlabel{validity:source1}{\text{(iv)}}\ &\neg\lozenge\!^- 0 \\
\customlabel{validity:source2}{\text{(v)}}\ &\lozenge^{\!+}\top\to 1 &
\customlabel{validity:mu-collapse1}{\text{(vi)}}\ &\mu p.\big(1\vee\lozenge^{\!+}Fp\big) \leftrightarrow 1 &
\customlabel{validity:mu-collapse2}{\text{(vii)}}\ &\mu p.\big(0\vee\lozenge^{\!-}Fp\big)\leftrightarrow 0. & &
\end{align*}

\end{proposition}
\begin{remark}
As a corollary of~\ref{validity:persistence}, $1\to G1$ follows by induction.
Validities~\ref{validity:source1} and~\ref{validity:source2} demonstrate the asymmetry between $\lozenge^+$ and $\lozenge^-$: an active reaction requires an active source, but may target an inactive vertex. 
Note that $\neg\lozenge^{\!+}0$ is not valid. 
This asymmetry has no analogue in standard modal logic.
Validity~\ref{validity:update} is a complete formal characterisation of the MR-network update rule. 
The disjuncts in~\ref{validity:convergence} are mutually exclusive.
Validity~\ref{validity:mu-collapse1} expresses that temporal reachability of `active' collapses to `currently active': since the source of any forward reaction must itself be active, the signal cannot originate from an inactive vertex.
The backward case~\ref{validity:mu-collapse2} is analogous: any vertex reached along backward reactions was necessarily active at an earlier time and hence remains active.
Note that our logic also includes the standard validities of hybrid logic (see e.g.~\cite{hybridlogic}).
\end{remark}
In Section~\ref{sec:other-models}, we outline how to interpret the syntax of $\mathbf{MRL}$ under the asynchronous update rule for MR-networks.

\section{Expressivity}\label{sec:expressivity}
We demonstrate the expressive power of $\mathbf{MRL}$ by formalising several properties of central importance in systems biology. For each, we discuss its biological motivation, give a precise $\mathbf{MRL}$-formalisation, and note any subtleties. We note that all the formulas in this section are readily interpreted under the asynchronous semantics introduced in Section~\ref{sec:other-models} -- hence capturing corresponding properties in a biologically more realistic scenario -- while the propositions and the examples are specific to the synchronous semantics studied here.

Throughout, we assume a fixed MR-network $\mathcal M=(V,R,M^\activation,M^\inhibition)$.

\begin{property}[Producibility]\label{property:producibility}
Given finite sets of nominals $X$ and $Y$, we define the formula ``$Y$ is producible from $X$'' as
$$\Prod(X,Y)\coloneq\left(\bigwedge_{x\in X} @_x\, 1\right) \to \left(\bigwedge_{y\in Y} @_y\, F 1 \right).$$
\end{property}
The formula $\Prod(X,Y)$ captures the situation in which initial activity of vertices $X$ guarantees that vertices $Y$ eventually become active. In chemical terms, this expresses the fact that an entity in $Y$ is necessarily among the products of $X$, if the system is allowed to interact long enough. We identify an MR-network condition sufficient for producibility.
\begin{definition}[Feedback closure]\label{def:feedback-closure}
Let $\mathcal M$ be an MR-network, and let $X\sse V$ be a subset of vertices. We define the {\em feedback closure} of $X$ as the subset $\overline X\sse V$ recursively constructed as follows. Base case: $X\sse\overline X$. Recursive case: for every $R$-path $(x_0,\dots,x_n)$ satisfying
\begin{itemize}
\item $x_0\in\overline X$,
\item for every $i=1,\dots,n$, there is an activation $z M^\activation (x_{i-1},x_i)$ such that either $z\in\overline X$ or $z=x_j$ for some $j\leq i-1$,
\item if $z M^\inhibition (x_{i-1},x_{i})$ is an inhibition for some reaction in the path, then $z=x_j$ for some $j\geq i$,
\end{itemize}
we stipulate that $x_n\in\overline X$.
\end{definition}
\begin{proposition}\label{prop:producible-feedback-closure}
Let $\mathcal M$ be an MR-network with $X,Y\sse V$ such that $Y\sse\overline X$. Then $\mathcal M\models\Prod(X,Y)$.
\end{proposition}
The simplest non-trivial instance of feedback closure is a path from $x$ to $y$ whose every reaction is activated by some vertex occurring before the reaction, and the only inhibitions come from vertices occurring after the reaction. We note that such paths exhibiting feedback inhibition have been studied by Savageau~\cite{savageau72} in the context of the power law approximation.

\noindent\begin{minipage}{.55\textwidth}
\begin{example}\label{ex:producibility}
We give an example of a path with feedback inhibition on the right. The feedback closure of $\{x\}$ is $\{x,y\}$, and we have $\mathcal M\models\Prod(x,y)$ and $\mathcal M\not\models\Prod(x,x_1)$.
\end{example}
\end{minipage}%
\begin{minipage}{.45\textwidth}
\vspace{-10pt}
\begin{equation*}
\scalebox{1}{\begin{tikzpicture}
	\begin{pgfonlayer}{nodelayer}
		\node [style=vertex] (0) at (-4, 0) {$x$};
		\node [style=vertex] (1) at (0, 0) {$x_1$};
		\node [style=none] (3) at (-2, 0.25) {};
		\node [style=vertex] (4) at (4, 0) {$y$};
		\node [style=none] (5) at (2, 0.25) {};
		\node [style=none] (7) at (-2, -0.25) {};
	\end{pgfonlayer}
	\begin{pgfonlayer}{edgelayer}
		\draw [style=reaction] (0) to (1);
		\draw [style=activation, bend left=90, looseness=1.25] (0) to (3.center);
		\draw [style=reaction] (1) to (4);
		\draw [style=activation, bend left=90, looseness=0.75] (0) to (5.center);
		\draw [style=inhibition, bend left=90, looseness=0.75] (4) to (7.center);
	\end{pgfonlayer}
\end{tikzpicture}}.
\end{equation*}
\end{minipage}

If we think of a subset $X$ as the biochemical entities that are initially present, finding the largest set $Y$ such that $\Prod(X,Y)$ is valid corresponds to {\em reaction prediction}, or exploring the full product space. Proposition~\ref{prop:producible-feedback-closure} implies that the feedback closure $\overline X$ gives a particularly well-behaved subset of the product space, which is computable using Definition~\ref{def:feedback-closure}: one starts with $X$, and iteratively adds all vertices reachable by paths that satisfy the specified conditions. The full product space is significantly more complicated to compute.

The opposite question to reaction prediction is that of {\em (retro)synthesis} or {\em production}: given a subset $Y$, can we find a (disjoint) subset $X$ such that $\Prod(X,Y)$? This has a much higher computational complexity than reaction prediction, as it is quantified over all subsets of an MR-network, thus scaling exponentially with the size of the network. In practice, the question of production, of course, also involves the design of the network.

The notion of producibility discussed above is a rather weak one: it merely records whether an entity is produced {\em at least once} during the computation. $\mathbf{MRL}$ can also capture a stronger notion.
\begin{property}[Sustained production]
Given a nominal $y$, we define the formula of {\em sustained production} of $y$ as $\Sust(y)\coloneq @_y\, F G \lozenge^{\!-}1$.
{\em Sustained producibility} on sets $X$ and $Y$ is then defined analogously to producibility (Property~\ref{property:producibility}):
$$\SustProd(X,Y)\coloneq\left(\bigwedge_{x\in X} @_x\, 1\right) \to \left(\bigwedge_{y\in Y} \Sust(y) \right).$$
\end{property}
The formula $\Sust(y)$ says that $y$ can be kept in a state where it is continuously receiving active input. Validity~\ref{validity:persistence} ($1 \to X 1$) implies that once a vertex becomes active it remains active. So the property of sustained production is not about the vertex value, but about whether the incoming reactions remain active.
 
In Example~\ref{ex:producibility}, we saw that $y$ is producible from $x$; however, $\SustProd(x,y)$ fails to be valid, as under the valuation $(x\mapsto 1, x_1\mapsto 0, y\mapsto 1)$ neither reaction will ever activate, so that there are no incoming active reactions to $y$. On the other hand, the formula $\SustProd(\{x,x_1\},y)$ is valid.

\begin{property}[Reachability]\label{property:reachability}
Given two nominals $x$ and $y$, we define the formula ``$y$ is reachable from $x$'' as
$$\Reach(x,y)\coloneq @_x\, \mu p.\left(y \vee F\lozenge^{\!+}p\right).$$
\end{property}
The formula $\Reach(x,y)$ is satisfied in $\mathcal M_i$ if and only if either $x=y$ or there is an $R$-path $(x=x_0,\dots,x_n=y)$ and a sequence of time instances $i_1,\dots,i_n$ satisfying $i\leq i_1\leq\cdots\leq i_n$ and $g_{i_j}\left(x_{j-1},x_j\right)=1$ for all $j=1,\dots,n$. Thus, $\Reach(x,y)$ generalises the usual notion of reachability within a graph (i.e.~the reflexive and transitive closure of the edge relation), while taking into account the fact that reactions can be absent or present. In Example~\ref{ex:mr-network2-det} we have $\mathcal M_0\models\Reach(x,y)$ but $\mathcal M_2\not\models\Reach(x,y)$, as the reaction $(x,y)$ will never reactivate. In Example~\ref{ex:producibility}, $\Reach(x,y)$ is satisfied if and only if $x$ is initially present and $y$ is initially absent; this suggests a more general link between reachability and producibility.
\begin{proposition}\label{prop:producibility-reachability}
$\mathcal M\models\Prod(x,y)$ implies $\mathcal M_0\models\Reach(x,y)$ with the valuation $f_0$ defined by $x\mapsto 1$ and $z\mapsto 0$ for $z\neq x$.
\end{proposition}
Note that the converse is not true, as there might be inactive inhibitions that prevent producibility. Thus, reachability can be thought of as a refined notion of producibility: it demands that all the reactions between $x$ and $y$ are activated in order, so that there is a material flow from $x$ to $y$.

The logic $\mathbf{MRL}$ is able to detect temporal cycles and loops in MR-networks, which we capture in the following two properties.
\begin{property}[Cycle]\label{property:cycle}
Given a nominal $x$, we define the formula for ``$x$ lies on a cycle'' as
$$\Cycle(x)\coloneq @_x\, F\lozenge^{\!+}\left(\mu p.\left(x \vee F\lozenge^{\!+}p\right)\right).$$
\end{property}
\begin{property}[Infinite forward path]\label{property:infinite-forward-path}
Given a nominal $x$, we define the formula for ``$x$ admits an infinite forward path'' as
$$\InfFwd(x)\coloneq @_x\, \nu p.\left(F\lozenge^{\!+}p\right).$$
\end{property}
Note that $\Cycle(x)$ is not equivalent to $\Reach(x,x)$, as the latter is valid ($x$ is always reachable from itself), whereas the former is satisfied if there is an $R$-path from $x$ to itself whose all reactions are activated in order. We, moreover, note that since all of our models are finite, satisfiability of $\InfFwd(x)$ implies presence of a cycle, though $x$ need not lie on it:
\begin{proposition}\label{prop:inf-path-to-cycle}
$\mathcal M_i\models\InfFwd(x)$ implies there is a $y\in V$ such that $\mathcal M_i\models\Reach(x,y)\wedge\Cycle(y)$.
\end{proposition}

\noindent\begin{minipage}{.6\textwidth}
\begin{example}\label{ex:cycle}
We give an example showing that the converse of Proposition~\ref{prop:inf-path-to-cycle} fails. Consider the catalytic cycle on the right with the initial valuation $i,y\mapsto 1$ and $0$ otherwise. We note that $\mathcal M_0\models\Reach(i,x)\wedge\Cycle(x)$, but $\mathcal M_0\not\models\InfFwd(i)$, as the reaction $(x,z)$ can be activated exactly once. In the network without the inhibition, we would indeed have that $\InfFwd(i)$ is satisfied.
\end{example}
\end{minipage}%
\begin{minipage}{.4\textwidth}
\vspace{-15pt}
\begin{equation*}
\scalebox{1}{\begin{tikzpicture}
	\begin{pgfonlayer}{nodelayer}
		\node [style=vertex] (0) at (4, 0) {$y$};
		\node [style=vertex] (1) at (0, 0) {$x$};
		\node [style=none] (3) at (2, 0.25) {};
		\node [style=vertex] (4) at (-4, 0) {$i$};
		\node [style=none] (5) at (-2, 0.25) {};
		\node [style=vertex] (6) at (2, -3.25) {$z$};
		\node [style=none] (7) at (0.75, -1.5) {};
		\node [style=none] (8) at (1, -2) {};
		\node [style=none] (9) at (3, -1.75) {};
	\end{pgfonlayer}
	\begin{pgfonlayer}{edgelayer}
		\draw [style=reaction] (0) to (1);
		\draw [style=activation, bend right=90, looseness=1.25] (0) to (3.center);
		\draw [style=activation, bend right=90, looseness=0.75] (0) to (5.center);
		\draw [style=reaction] (4) to (1);
		\draw [style=reaction] (1) to (6);
		\draw [style=reaction] (6) to (0);
		\draw [style=activation, bend right=90, looseness=1.50] (1) to (7.center);
		\draw [style=inhibition, bend left=90, looseness=1.50] (6) to (8.center);
		\draw [style=activation, bend right=90, looseness=1.50] (6) to (9.center);
	\end{pgfonlayer}
\end{tikzpicture}}.
\end{equation*}
\end{minipage}

The presence of cycles becomes more interesting under semantics where vertices can deactivate, such as the asynchronous semantics of Section~\ref{sec:other-models}.

We list further properties expressible in $\mathbf{MRL}$ in the table below, partially inspired by the properties in~\cite{biocham2005}:
\begin{center}
\begin{tabular}{c | c}
$\mathbf{MRL}$ formula & Meaning \\ \hline
$@_x\, \mu p.\left(z \vee F\lozenge^{\!+}p\right)\wedge @_z\, \mu q.\left(y \vee F\lozenge^{\!+}q\right)$ & $y$ is reachable from $x$ via $z$ \\ \hline
$\mu p.\left(\varphi\vee (\psi\wedge Xp) \right)$ & $\psi$ holds until $\varphi$ holds (weak until) \\ \hline
$\mu p.\left(@_x\, 1\vee (\neg(@_y\, 1)\wedge Xp) \right)$ & $x$ is producible without producing $y$ \\ \hline
$@_x\, \left(1\to\nu p.\left(1\wedge Xp\right)\right)$ & $x$ is a steady state \\ \hline
$@_x\, (1\to G1)$ & $x$ is a permanent state \\ \hline
$\bigwedge_{a\in A}@_a\, G\square^{\!+}\left(\bigvee_{a'\in A}a'\right)$ & $A$ is a sink set \\ \hline
$\bigwedge_{b\in B}@_b\, \mu p.\left(\bigvee_{a\in A}a\vee \left(F\lozenge^{\!+}p\wedge\square^{\!+}p\right)\right)$ & $B$ flows into $A$
\end{tabular}
\end{center}
The properties expressing steady and permanent states are meaningful only in a semantics where vertices can deactivate, as under current semantics they are validities.

The last two properties can be used to define {\em attractors}. Let us write $\Sink(A)$ for ``$A$ is a sink set'' and $\Flow(B,A)$ for ``$B$ flows into $A$''. $\Sink(A)$ is satisfied if there are no outgoing active reactions from $A$ -- once matter flows into $A$, it is trapped in $A$. In Example~\ref{ex:producibility}, the formulas $\Sink(y)$ and $\Sink(x_1,y)$ are evidently valid; less evidently, the formula $XX\Sink(x)$ is also valid. $\Flow(B,A)$ is satisfied if from every vertex in $B$ some vertex in $A$ is reachable, and all paths that start from $B$ end up in $A$.
\begin{property}[Attractors]\label{property:attractors}
For sets of nominals $A$ and $B$, we define the formula ``$A$ is a preattractor of $B$'' as $\PreAttr(A,B)\coloneq\Sink(A)\wedge\Flow(B,A)$.
We then define ``$A$ is the attractor of $B$'' as
$$\Attr(A,B)\coloneq\PreAttr(A,B)\bigwedge_{A'\subsetneq A}\neg\PreAttr(A',B).$$
\end{property}
The attractor of $B$ is thus the minimal sink set $A$ into which $B$ flows: every proper subset of $A$ is either not a sink set or $B$ does not flow into it. In Example~\ref{ex:cycle}, we have $\mathcal M_0\models\Flow(V,z)$, but $z$ cannot be a (pre)attractor as it is not a sink. Instead, we have $\mathcal M_0\models\Attr(\{x,y,z\},V)$ and $\mathcal M_2\models\Attr(x,V)$.

\section{Evaluation games and bisimulation}\label{sec:bisimulation}

Here we characterise satisfiability of $\mathbf{MRL}$-formulas in MR-models in two alternative ways: {\em evaluation games} give an algorithm for model-checking, while invariance under {\em bisimulation} is used to obtain inexpressibility results.

\subsection{Evaluation games}

The Tarski-style semantics of Section~\ref{sec:formal-language} defines truth via set-theoretic fixed-points. The evaluation game provides an equivalent game-theoretic characterisation in which truth is witnessed by a concrete strategy.
Existence of a winning strategy (Theorem~\ref{thm:adequacy}) gives an algorithm to determine whether a property is satisfied in an MR-network, thus giving a complexity bound for model-checking (Corollary~\ref{cor:model-checking}).
To demonstrate that a vertex satisfies a formula, Verifier must exhibit the relevant neighbours, choose the right future time steps, and unfold fixed-points until an atom is reached. 
In MR-networks, this has a direct biological consequence: a play of the game traces a signal propagation path through the network across time, constrained at each step by active reactions.

Every formula of $\mathbf{MRL}$ can be brought into
\emph{negation normal form} (NNF) by pushing negations inward using De Morgan duality, the dual modalities defined in Section~\ref{sec:formal-language}, the self-duality $\neg X\varphi \equiv X\neg\varphi$ (since the temporal successor is deterministic), and the atom dualities $\neg 0 \equiv 1$, $\neg 1 \equiv 0$. 
For example, the NNF of $\neg\Diamond^+\neg 0$ is $\Box^+ 0$, and the NNF of $\neg F(1 \land \Diamond^+ 0)$ is $G(0 \vee \Box^+ 1)$.
We assume throughout this subsection that formulas are in NNF.
 
Let $\varphi$ be an $\mathbf{MRL}$ sentence in NNF.
The \emph{evaluation game} $\mathcal E(\varphi,\mathcal M_i,x)$ is played between two players, \textbf{Verifier} ($\exists$) and \textbf{Falsifier} ($\forall$), starting from position $(\varphi,(x,i))$.
A \emph{position} is a pair $(\psi,(y,j))$ with $\psi\in\mathrm{cl}(\varphi)$ and $(y,j)\in V\times\N$, where $\mathrm{cl}(\varphi)$ denotes the \emph{closure} of $\varphi$, i.e.\ the set of subformulas of $\varphi$, closed under unfolding of
fixed-point variables.
\vspace{-1em}
\begin{table}[h]
    \centering
\renewcommand{\arraystretch}{1.35}
\scalebox{0.88}{
\begin{tabular}{ccc}
\toprule
\textbf{Position} & \textbf{Move} & \textbf{Next position} \\
\midrule
$(\psi_1\vee\psi_2,\;(y,j))$ & $\exists$ picks $k\in\{1,2\}$ &
  $(\psi_k,(y,j))$ \\
$(\psi_1\wedge\psi_2,\;(y,j))$ & $\forall$ picks $k\in\{1,2\}$ &
  $(\psi_k,(y,j))$ \\
$(\lozenge^+\!\psi,\;(y,j))$ & $\exists$ picks $z$ s.t.~$yRz$, $g_j(y,z)=1$ &
  $(\psi,(z,j))$ \\
$(\square^+\!\psi,\;(y,j))$ & $\forall$ picks $z$ s.t.~$yRz$, $g_j(y,z)=1$ &
  $(\psi,(z,j))$ \\
$(\lozenge^-\!\psi,\;(y,j))$ & $\exists$ picks $z$ s.t.~$zRy$, $g_j(z,y)=1$ &
  $(\psi,(z,j))$ \\
$(\square^-\!\psi,\;(y,j))$ & $\forall$ picks $z$ s.t.~$zRy$, $g_j(z,y)=1$ &
  $(\psi,(z,j))$ \\
$(X\psi,\;(y,j))$ & \;\;\; $\mathrel{\rule[0.5ex]{1em}{0.4pt}}$ &
  $(\psi,(y,j\!+\!1))$ \\
$(F\psi,\;(y,j))$ & $\exists$ picks $\ell\geq j$ &
  $(\psi,(y,\ell))$ \\
$(G\psi,\;(y,j))$ & $\forall$ picks $\ell\geq j$ &
  $(\psi,(y,\ell))$ \\
$(@_a\psi,\;(y,j))$ & \;\;\; $\mathrel{\rule[0.5ex]{1em}{0.4pt}}$ &
  $(\psi,(a,j))$ \\
$(\mu p.\psi,\;(y,j))$ & \;\;\; $\mathrel{\rule[0.5ex]{1em}{0.4pt}}$ &
  $(\psi,(y,j))$ \\
$(\nu p.\psi,\;(y,j))$ & \;\;\; $\mathrel{\rule[0.5ex]{1em}{0.4pt}}$ &
  $(\psi,(y,j))$ \\
$(p,\;(y,j))$ for $p$ bound by $\sigma$ & \;\;\; $\mathrel{\rule[0.5ex]{1em}{0.4pt}}$ &
  $(\sigma p.\psi,(y,j))$ \\
\bottomrule
\end{tabular}}
 \vspace{1em}
  \caption{Table of moves\label{fig:movestable}}
\end{table}

\paragraph{Moves and winning condition.}
From a position $(\psi,(y,j))$ the game proceeds according to the
outermost connective of $\psi$ as depicted in Table~\ref{fig:movestable}.

If the active player at a $\lozenge^{(\cdot)}$ or $\square^{(\cdot)}$ position
has no legal move (i.e.\ no active edge in the required direction),
that player loses immediately.
A play at an atom $a\in\{0,1\}$ terminates:
$\exists$ wins if $f_j(y)=a$; $\forall$ wins otherwise.
Similarly, at a nominal $a$ (resp.\ $\neg a$), $\exists$ wins if $y=a$
(resp.\ $y\neq a$). A play may be infinite if it passes through fixed-point positions
$\mu p.\psi$ or $\nu p.\psi$ infinitely often.
To each fixed-point subformula $\sigma p.\psi$ of $\varphi$ (where $\sigma\in\{\mu,\nu\}$) we assign a \emph{priority} $\Omega(\sigma p.\psi)\in\N$, such that:
\begin{enumerate}
    \item $\Omega(\mu p.\psi)$ is odd and $\Omega(\nu p.\psi)$ is even;
    \item if $\sigma_1 p_1.\psi_1$ is a proper subformula of $\sigma_2 p_2.\psi_2$, then $\Omega(\sigma_1 p_1.\psi_1)>\Omega(\sigma_2 p_2.\psi_2)$.
\end{enumerate}
An infinite play is won by $\exists$ if and only if the highest priority visited infinitely often is \emph{even}.
Thus, infinite looping through a $\nu$-variable is favourable to $\exists$ (the property persists), while infinite looping through a $\mu$-variable is favourable to $\forall$ (the least fixed-point is never witnessed).

\noindent\begin{minipage}{.37\textwidth}
\begin{example}\label{ex:eval-game}
Consider the MR-network of Example~\ref{ex:producibility} with the initial valuation $x \mapsto 1$, $x_1 \mapsto 0$, $y \mapsto 0$. We demonstrate the evaluation game on $\mathsf{Reach}(x, y) = @_x \mu p. (y \vee F\Diamond^+ p)$. 
The game tree is depicted on the right.
Note that all moves in this game are verifier moves.
 \end{example}
\end{minipage}\;\;\;
\begin{minipage}{0.63\textwidth}
\vspace{-20pt}
\begin{equation*}
\scalebox{0.75}{%
  \begin{tikzpicture}[
  every node/.style={font=\small},
  box/.style={draw, rounded corners=3pt, fill=gray!5, minimum width=2.8cm, minimum height=0.7cm, inner sep=4pt, align=center},
  lose/.style={box, fill=gray!10, draw=gray!60},
  win/.style={box, fill=green!8, draw=green!50!black},
  lbl/.style={font=\small\itshape, text=black!70},
  ell/.style={font=\small, text=black!60},
  >=stealth,
  x=1.2cm, y=1.2cm
]

\node[box] (r0) at (0,0) {$y \vee F\Diamond^+ p,\; (x,0)$};

\node[lose] (l1) at (-3,-2) {$x \neq y$\\$\exists$ loses};
\node[box]  (d1) at (3,-2) {$\Diamond^+ p,\; (x,0)$};

\draw[->] (r0) -- (l1) node[lbl, midway,  left] {$\exists$ picks $y$\;\;};
\draw[->] (r0) -- (d1) node[lbl, midway,  right] {\;$\exists$ picks $F\Diamond^+ p$};
\node[ell] at (6,-2) {$\ell = 0$};

\node[box] (r2) at (3,-4) {$y \vee F\Diamond^+ p,\; (x_1,0)$};
\draw[->] (d1) -- (r2) node[lbl, midway, right] {$\exists$ picks $x_1$};

\node[lose] (l3) at (0,-6) {$x_1 \neq y$\\$\exists$ loses};
\node[box]  (d3) at (6,-6) {$\Diamond^+ p,\; (x_1,1)$};

\draw[->] (r2) -- (l3) node[lbl, midway, left ] {$\exists$ picks $y$\;\;};
\draw[->] (r2) -- (d3) node[lbl, midway, right ] {\;$\exists$ picks $F\Diamond^+ p$};
\node[ell] at (9,-6) {$\ell = 1$};

\node[box] (r4) at (6,-8) {$y \vee F\Diamond^+ p,\; (y,1)$};
\draw[->] (d3) -- (r4) node[lbl, midway, right] {$\exists$ picks $y$};

\node[win] (w) at (6,-10) {$y = y$\\$\exists$ wins};
\draw[->] (r4) -- (w) node[lbl, midway, right] {$\exists$ picks $y$};

\end{tikzpicture}}
\end{equation*}
\end{minipage}

\begin{theorem}[Adequacy]\label{thm:adequacy}
  Let $\varphi$ be an $\mathbf{MRL}$ sentence in NNF. 
  Then 
  $$\mathcal M_i,x\models\varphi\iff \text{$\exists$ has a winning strategy in $\mathcal E(\varphi,\mathcal M_i,x)$.}
 $$
\end{theorem}
\begin{corollary}[Model-checking complexity]\label{cor:model-checking}
    For finite MR-networks, model-checking $\mathbf{MRL}$ reduces to solving a parity game and is hence decidable in $\mathbf{NP}\cap\mathbf{coNP}$.
    More concretely, checking $\mathcal M_i,x\models\varphi$ can be done in time $O(|V|^d\cdot|\varphi|^d)$ where $d$ is the alternation depth of~$\varphi$.
\end{corollary}
\subsection{Bisimulation}
Bisimulation is a relation between two MR-models whose existence guarantees that related vertices satisfy the same $\mathbf{MRL}$ sentences.

\begin{definition}[Bisimulation]\label{def:bisimulation}
Let $(\mathcal M,f_0)$ and $(\mathcal N,f'_0)$ be MR-models with vertex sets $V$ and $W$, edge relations $R$ and $S$, and induced state sequences $(f_i,g_i)_{i\in\N}$ and $(f'_i,h_i)_{i\in\N}$, respectively.
An \emph{MRL-bisimulation} is a relation $Z\subseteq (V\times\N)\times(W\times\N)$ satisfying the following conditions whenever $(v,k)\mathrel{Z}(w,\ell)$:

\begin{enumerate}[label={\rm{(\arabic*)}}]
    \item  $f_k(v)=f'_\ell(w)$.

\item  $(v,k+1)\mathrel{Z}(w,\ell+1)$.

    \item $(a^{\mathcal M},k)\mathrel{Z}(a^{\mathcal N},\ell)$, for each nominal $a\in\mathbf{Nom}$.

    \item \textbf{($\lozenge^+$-forth)}
    If $vRv'$ and $g_k(v,v')=1$, then there exists $w'\in W$
    with $wSw'$, $h_\ell(w,w')=1$, and $(v',k)\mathrel{Z}(w',\ell)$.

    \item \textbf{($\lozenge^+$-back)}
    If $wSw'$ and $h_\ell(w,w')=1$, then there exists $v'\in V$
    with $vRv'$, $g_k(v,v')=1$, and $(v',k)\mathrel{Z}(w',\ell)$.

    \item \textbf{($\lozenge^-$-forth)}
    If $v'Rv$ and $g_k(v',v)=1$, then there exists $w'\in W$
    with $w'Sw$, $h_\ell(w',w)=1$, and $(v',k)\mathrel{Z}(w',\ell)$.

    \item \textbf{($\lozenge^-$-back)}
    If $w'Sw$ and $h_\ell(w',w)=1$, then there exists $v'\in V$
    with $v'Rv$, $g_k(v',v)=1$, and $(v',k)\mathrel{Z}(w',\ell)$.

\end{enumerate}
We write $(\mathcal M_k,v)\sim(\mathcal N_\ell,w)$
if there exists an MRL-bisimulation $Z$ such that $(v,k)\mathrel{Z}(w,\ell)$.
\end{definition}

\begin{theorem}[Bisimulation invariance]\label{thm:bisim-invariance}
  If $(\mathcal M_k,v)\sim(\mathcal N_\ell,w)$,
  then for every $\mathbf{MRL}$ sentence $\varphi$ we have
 $\mathcal M_k,v\models\varphi
    \iff
    \mathcal N_\ell,w\models\varphi$.
\end{theorem}
The properties defined in Section~\ref{sec:expressivity} demonstrate that $\mathbf{MRL}$ can detect the \emph{existence} of active paths, cycles, and temporal reachability.
Here we use bisimulation invariance to show that $\mathbf{MRL}$ cannot distinguish the \emph{number}
of active reactions incident to a vertex.

\begin{proposition}\label{prop:indis}
There is no $\mathbf{MRL}$ sentence $\varphi$ such that
for all MR-models $(\mathcal M,f_0)$ one has
$\mathcal M_i,x\models\varphi
    \iff
    \big|\{y\in V: yRx, g_i(y,x)=1\}\big|=2$.
\end{proposition}
A second source of indistinguishability is structural: two MR-networks with distinct modulation structures but identical induced dynamics $(f_i, g_i)_{i \in \mathbb{N}}$ satisfy the same $\mathbf{MRL}$ sentences at every vertex and time, as the satisfaction relation depends only on $f_i$ and $g_i$. 
Thus, $\mathbf{MRL}$ is blind both to multiplicities of active reactions and to the particular modulation mechanism.

\section{Other models of computation}\label{sec:other-models}

Here we briefly discuss how MR-networks and their synchronous semantics connect to other models of computation. We do this by encoding MR-networks as Boolean networks, as well as by describing two possibilities for more elaborate semantics: {\em asynchronous Boolean semantics}, resulting in a nondeterministic computation, and {\em discrete quantitative dynamics}, which is deterministic.

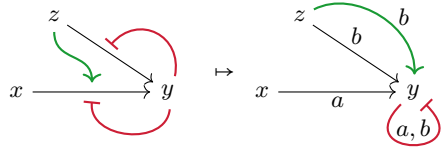
\begin{figure}[h]
\centering
\scalebox{1}{\begin{tikzpicture}
	\begin{pgfonlayer}{nodelayer}
		\node [style=vertex] (0) at (-5.5, -0.5) {$x$};
		\node [style=vertex] (1) at (-1.5, -0.5) {$y$};
		\node [style=vertex] (2) at (-4.5, 1.5) {$z$};
		\node [style=none] (3) at (-3.5, -0.25) {};
		\node [style=none] (4) at (-3, 0.75) {};
		\node [style=none] (5) at (-3.5, -0.75) {};
		\node [style=none] (6) at (0, 0) {$\mapsto$};
		\node [style=vertex] (7) at (1, -0.5) {$x$};
		\node [style=vertex] (8) at (5, -0.5) {$y$};
		\node [style=vertex] (9) at (2, 1.5) {$z$};
		\node [style=none] (10) at (3, -0.75) {$a$};
		\node [style=none] (11) at (3.5, 1) {$b$};
		\node [style=none] (12) at (4.75, 1.5) {$b$};
		\node [style=none] (13) at (5, -1.5) {$a,b$};
	\end{pgfonlayer}
	\begin{pgfonlayer}{edgelayer}
		\draw [style=reaction] (0) to (1);
		\draw [style=activation, in=90, out=-90, looseness=1.25] (2) to (3.center);
		\draw [style=reaction] (2) to (1);
		\draw [style=inhibition, bend right=75, looseness=1.50] (1) to (4.center);
		\draw [style=inhibition, bend left=90, looseness=1.25] (1) to (5.center);
		\draw [style=reaction] (7) to (8);
		\draw [style=reaction] (9) to (8);
		\draw [style=inhibition, in=-45, out=-135, loop] (8) to ();
		\draw [style=activation, bend left=60] (9) to (8);
	\end{pgfonlayer}
\end{tikzpicture}}
\caption{An encoding of an MR-network as a Boolean network.\label{fig:mr-to-regulatory}}
\AltTextCMSB{An encoding of an MR-network as a Boolean network}
\end{figure}

\paragraph{Translation to Boolean networks.}
Every MR-network can be encoded as a regulatory graph with the same vertices and three binary relations: {\em reaction}, {\em activation} and {\em inhibition} relations, as well as a label for each edge in the graph. We give an example translation from an MR-network to the corresponding regulatory graph in Fig.~\ref{fig:mr-to-regulatory}.
The translation thus assigns a unique label to each reaction, and shifts the target of a modulation to the target of the reaction it modulates, while adding the label of the reaction to the set of labels of the modulation. In the resulting regulatory graph, each modulation is, therefore, labelled with a non-empty subset of the reaction labels. The Boolean update function for a vertex $v$ is then defined analogously to MR-networks: $v$ activates if and only if $v$ is already active, or there is a reaction $(u,v)$ with label $a$ such that $u$ is active, there is at least one activation with label $a$ and no inhibitions with label $a$. While this translation is a faithful representation of MR-networks (in the sense of being an injective map from MR-networks to Boolean networks), the conceptual intuition of reactions appearing and disappearing as the computation proceeds is lost. 
For example, while it is still possible to define the semantics of the reachability modalities $\lozenge^{\!+}$ and $\lozenge^{\!-}$, the definitions become rather cumbersome. 
Moreover, the translation is tied to the synchronous semantics: e.g.~the asynchronous semantics we shall discuss next is not equivalent to the usual asynchronous update scheme of Boolean networks, as (in)activity of a reaction is part of a state, and can thus be updated without any of the vertices being updated. Nonetheless, we find it important to point out the existence of such translation, as it can help with implementation as well as with establishing connections to well-understood models of computation, such as Petri nets~\cite{compositional-account2024}.

\paragraph{Asynchronous Boolean semantics.}
The asynchronous Boolean semantics is given as follows, which is a restricted version of the semantics introduced in~\cite{rougny2016}:
\begin{itemize}
\item a reaction can go from inactive to active if its source is active, at least one of its activators is active, and none of its inhibitors are active (that is, precisely when the update condition~\eqref{eq:reaction-update} is satisfied),
\item a reaction can go from active to inactive if the activation condition above is no longer satisfied, or if its target is active,
\item a vertex can go from inactive to active if there is at least one active incoming reaction (that is, precisely when the update condition~\eqref{eq:vertex-update} is satisfied),
\item a vertex can go from active to inactive if there is an active outgoing reaction.
\end{itemize}
The logic $\mathbf{MRL}$ can be interpreted over the asynchronous semantics: the assignment function now takes the form $\sigma : \mathbf{Var}\to\mathcal{P}(V\times (V\sqcup R)^*)$, i.e.~a proposition is interpreted as a pair of a vertex and a finite sequence of vertices and reactions. The set of such sequences $(V\sqcup R)^*$ plays the role of time: each vertex and reaction in the sequence attempts to change its state according to the rules above. The next step modality $X$ now becomes a {\em one-step possibility modality}: $X\varphi$ holds if and only if there exists a single vertex or a reaction such that $\varphi$ holds once the update is applied. Similarly, $F\varphi$ holds if there is a sequence in $(V\sqcup R)^*$ such that $\varphi$ holds once all updates are applied. We also need to change the notion of a model to include the initial state (rather than merely the initial valuation), as the state is no longer uniquely determined by the valuation. After these modifications, the $\mathbf{MRL}$ formulas are interpreted as in Section~\ref{sec:formal-language}. The set of validities is, of course, now distinct, although we note that the modalities $F$ and $G$ can still be obtained via fixed-point operators as before.

\paragraph{Discrete quantitative dynamics.}
The state of discrete quantitative dynamics is given by assigning a nonnegative integer to each vertex and each reaction. The interpretation for vertices is (discretised) concentration of a given entity, while for reactions the integer $n$ tells the flow rate (cf.~\cite{integer-hyperflows2019}). At each step, each reaction attempts to transport $n$ units from its source to its target. We additionally assume that each reaction and each modulation is labelled by a positive real number, which represents the strength of the modulation and global reaction rate. In order to have interesting behaviour, we consider {\em open} MR-networks, whose vertices come with specified subsets of {\em inputs} and {\em outputs}. The input and the output sets are assumed to be disjoint, the inputs are assumed to have no incoming reactions, and the outputs are assumed to have no outgoing reactions. The input and output vertices correspond to the {\em empty set} notation of SBGN-PD for sources and sinks. Semantically, to each input we assign a function $\omega\rightarrow\N$, i.e.~a {\em stream} of nonnegative integers, specifying the influx of new matter at each time step. Correspondingly, the update rule for vertices induces a stream for each output. Hence an open MR-network can be thought of as a function $\left(\N^\omega\right)^n\rightarrow\left(\N^\omega\right)^m$, where $n$ is the number of inputs and $m$ is the number of outputs, giving an interesting connection to the theory of {\em signal flow graphs}~\cite{shannon42,rutten-tutorial,survey-signal-flow}. Let us denote the labelling function for reactions and modulations by $\ell:R\sqcup M^\activation\sqcup M^\inhibition\rightarrow\R_+$. Given an initial valuation $f_0:V\Setminus I\rightarrow\N$ on the non-input vertices and a stream $s_v:\omega\rightarrow\N$ for each input vertex $v\in I$, we first extend the valuation to the initial state $f_0:V\rightarrow\N$ by defining $f_0(v)\coloneq s_v(0)$ on the input vertices $v\in I$. The one-step update rule $f_i\mapsto f_{i+1}$ is then given as follows:
\begin{itemize}
\item for each reaction $r$, the {\em potential flow} is given by
$$\PF_i(r)\coloneq\left\lfloor\ell(r)\cdot\left(\sum_{(v,r)\in M^\activation}\ell(v,r)\cdot f_i(v) - \sum_{(v,r)\in M^\inhibition}\ell(v,r)\cdot f_i(v)\right)\right\rfloor$$
if the difference between the sums is nonnegative, and by $\PF_i(r)\coloneq 0$ otherwise,
\item for each vertex $v$, the {\em potential outflow} is given by
$$\POF_i(v)\coloneq\sum_{(v,u)\in R}\PF_i(v,u);$$
if $\POF_i(v)\leq f_i(v)$, then for each outgoing reaction the {\em realised flow} is given by its potential flow: $\RF_i(v,u)\coloneq\PF_i(v,u)$; if $\POF_i(v) > f_i(v)$, then for each outgoing reaction the realised flow is a fraction of the current state weighted by its potential flow: $\RF_i(v,u)\coloneq\left\lfloor\frac{\PF_i(v,u)}{\POF_i(v)}\cdot f_i(v)\right\rfloor$,
\item for a non-input and non-output vertex $v\in V\Setminus (I\cup O)$, the updated state is given by
$$f_{i+1}(v)\coloneq f_i(v) - \sum_{(v,u)\in R}\RF_i(v,u) + \sum_{(u,v)\in R}\RF_i(u,v);$$
for an input vertex $v\in I$, the updated state is given as above, except that the last sum is replaced by the stream value $s_v(i+1)$; for an output vertex $v\in O$, the updated state is simply given by the last sum above, i.e.~by the sum of all the realised flows into it.
\end{itemize}

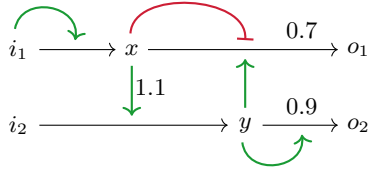
\begin{figure}[h]
\centering
\scalebox{1}{\begin{tikzpicture}
	\begin{pgfonlayer}{nodelayer}
		\node [style=vertex] (0) at (-4.5, 1) {$i_1$};
		\node [style=vertex] (1) at (-4.5, -1) {$i_2$};
		\node [style=vertex] (2) at (-1.5, 1) {$x$};
		\node [style=vertex] (3) at (4.5, 1) {$o_1$};
		\node [style=vertex] (4) at (4.5, -1) {$o_2$};
		\node [style=vertex] (5) at (1.5, -1) {$y$};
		\node [style=none] (6) at (1.5, 1.25) {};
		\node [style=none] (7) at (1.5, 0.75) {};
		\node [style=none] (8) at (-1.5, -0.75) {};
		\node [style=none] (9) at (-3, 1.25) {};
		\node [style=none] (10) at (3, -1.25) {};
		\node [style=none] (12) at (-1, 0) {$1.1$};
		\node [style=none] (14) at (3, -0.5) {$0.9$};
		\node [style=none] (19) at (3, 1.5) {$0.7$};
	\end{pgfonlayer}
	\begin{pgfonlayer}{edgelayer}
		\draw [style=reaction] (0) to (2);
		\draw [style=reaction] (2) to (3);
		\draw [style=reaction] (1) to (5);
		\draw [style=reaction] (5) to (4);
		\draw [style=activation] (2) to (8.center);
		\draw [style=activation] (5) to (7.center);
		\draw [style=inhibition, bend left=90] (2) to (6.center);
		\draw [style=activation, bend left=90, looseness=1.50] (0) to (9.center);
		\draw [style=activation, bend right=90, looseness=1.50] (5) to (10.center);
	\end{pgfonlayer}
\end{tikzpicture}}
\caption{An open MR-network with labels.\label{fig:open-mr-labels}}
\AltTextCMSB{An open MR-network with labels}
\end{figure}
\begin{figure}[h]
\centering
\includegraphics[scale=0.25]{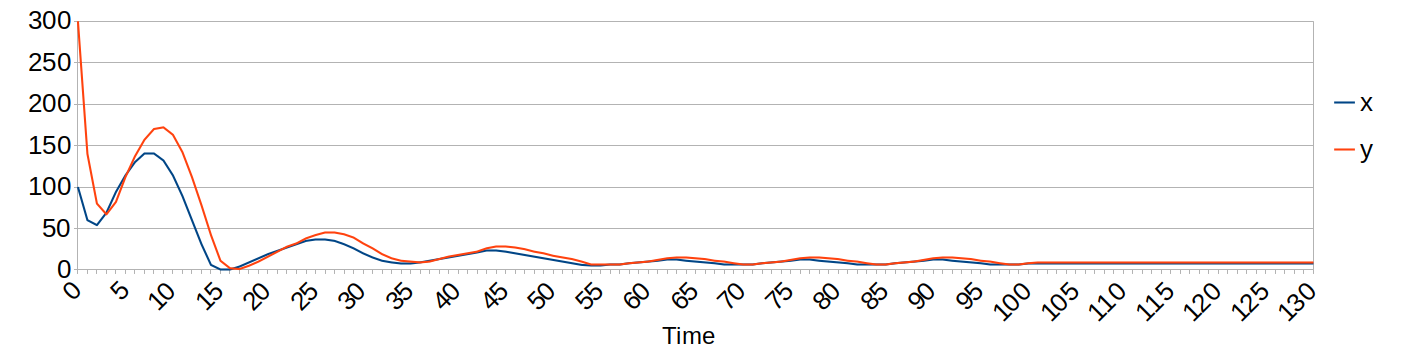}
\caption{An example computation of an open labelled MR-network.\label{fig:example-computation}}
\AltTextCMSB{An example computation of an open labelled MR-network}
\end{figure}

Consider the open MR-network with labels in Fig.~\ref{fig:open-mr-labels}, which is the network from Fig.~\ref{fig:reg-graph}(B) with added labels. We adopt the convention that an unlabelled edge stands for label $1$, i.e.~an unweighted reaction or modulation. The input vertices are $\{i_1,i_2\}$ and the output vertices are $\{o_1,o_2\}$. Let us define the initial valuation on the non-input vertices as $f_0(x)=100$, $f_0(y)=300$ and $f_0(o_1)=f_0(o_2)=0$, and the streams $s_1(n)\coloneq\left\lfloor\frac{100}{n+1}\right\rfloor$ and $s_2(n)\coloneq 150$ corresponding to the input vertices $i_1$ and $i_2$. The structure of the network implies that the stream $s_1$ is increasing the entities $x$, so that we keep increasing $x$ by a decreasing amount, until after time $n=100$ there is no new inflow of $x$. 
The stream $s_2$ is so chosen in order to have an essentially unlimited supply, so that the only factor on which the production of entities $y$ depends is the number of entities $x$. We plot the values of entities $x$ and $y$ for the first $130$ steps of the computation in Fig.~\ref{fig:example-computation}. We observe damped oscillatory behaviour that reaches a fixed-point once the input from $i_1$ stops, which is exactly the behaviour of the continuous time dynamical system obtained for this network by Voit~\cite{voit13}.

\section{Conclusion and future work}\label{sec:conclusion}

We have introduced MR-networks, their synchronous computation, as well as their logic $\mathbf{MRL}$.
Three perspectives on the satisfiability of $\mathbf{MRL}$-formulas allow formulating {\em expressibility}, {\em model-checking} and {\em inexpressibility} of biological properties in $\mathbf{MRL}$ with respect to the synchronous semantics. 
Our usage of the fixed-point modalities streamlines and unifies many properties studied in the literature on a case-by-case basis. 
While the synchronous update rule is inadequate for most biological applications, we believe that the modular approach taken here -- separating networks, computation, formal language and its interpretation -- lays the groundwork for more general study of MR-networks and their use in applications. 
For example, the properties in Section~\ref{sec:expressivity} can be readily interpreted in the asynchronous semantics described in Section~\ref{sec:other-models}: what changes are the precise details of models and networks captured by a formula -- the overarching biological concept remains the same.
In a forthcoming article, we will extend all the developments here -- expressibility, model-checking and inexpressibility -- to the asynchronous semantics.

Beyond the Boolean case, we have described a quantitative discrete semantics in Section~\ref{sec:other-models}. An interesting future development would be to give a logic for this semantics. The atomic case would need to be extended to capture the concentration being above (or below) a given threshold, while the other properties of $\mathbf{MRL}$ would be expected to carry over to the quantitative case.

In order to represent realistic reactions with multiple reactants and products (usually represented using multisets~\cite{inferring-reaction-systems2015,abstract-simulation2022}), one has to extend MR-networks from graphs to hypergraphs: the relation $R$ would no longer be binary, but would specify a set of sources and a set of targets~\cite{integer-hyperflows2019}. Another step closer to a more complete description of biochemical processes will be accounting for other kinds of modulations present in SBGN-PD~\cite{sbgn-pd19}.

In the direction of quantitative semantics, it would be interesting to define composition of open MR-networks, following e.g.~the approach of~\cite{chaudhuri2026} for SBGN-PD diagrams. This would provide a systematic way of building larger MR-networks from smaller ones.

Lastly, for comparing different semantics, it is important to connect MR-networks to model reduction and abstract interpretation techniques~\cite{fages2008,static-analysis2012,influence-compared2018,model-reduction2026}.

\begin{credits}
\subsubsection{\ackname} 
\begin{sloppypar}
    Yo\`av Montacute is supported by ACT-X, Grant No.\ \mbox{JPMJAX24CR}, JST, Japan; and by ASPIRE, Grant No.\ JPMJAP2301, JST, Japan.
Leo Lobski is partially supported by ARIA Safeguarded AI programme, UK.
\end{sloppypar}

\subsubsection{\discintname}
The authors have no competing interests to declare that are
relevant to the content of this article. 
\end{credits}
\bibliographystyle{splncs04}
\bibliography{bibliography}

\end{document}